\documentclass[aip,reprint]{revtex4-1}
\usepackage{graphicx}
\usepackage{CJK}
\usepackage{dcolumn}
\usepackage{bm}

\usepackage[utf8]{inputenc}
\usepackage[T1]{fontenc}
\usepackage{etoolbox}
\usepackage{dsfont}
\usepackage{amsmath,amsthm,amssymb}
\usepackage{mathtools}
\usepackage{color}
\usepackage{soul}

\newcommand{\V}[1]{\boldsymbol{#1}} 
\newcommand{\M}[1]{\boldsymbol{#1}} 
\newcommand{\abs}[1]{\left|#1\right|} 

\newcommand{\grad}{\M{\nabla}} 

\makeatother
\begin{document}

\preprint{AIP/123-QED}

\title[]{Mechanisms of electrostatic interactions between two charged dielectric spheres inside a polarizable medium: An effective-dipole analysis}

\author{Yanyu Duan}
\affiliation{Thrust of Advanced Materials, and Guangzhou Municipal Key Laboratory of Materials Informatics, Function Hub, The Hong Kong University of Science and Technology (Guangzhou), China.}
 
\author{Zecheng Gan}
\email{zechenggan@hkust-gz.edu.cn; \\
       zechenggan@ust.hk}
\affiliation{Thrust of Advanced Materials, and Guangzhou Municipal Key Laboratory of Materials Informatics, Function Hub, The Hong Kong University of Science and Technology (Guangzhou), China.}
\affiliation{Department of Mathematics, The Hong Kong University of Science and Technology, Hong Kong, China.}

\author{Ho-Kei Chan}
\email{hokeichan@hit.edu.cn; \\
       epkeiyeah@yahoo.com.hk}
\affiliation{School of Science, Harbin Institute of Technology (Shenzhen), China.}

\date{\today}

\begin{abstract}
The mechanisms of electrostatic interactions between two charged dielectric spheres inside a polarizable medium have been investigated, in terms of hypothetical effective dipoles that depict how the positive and negative charge in each particle are separated. Our findings, which revealed that it is possible for polarization-induced opposite-charge repulsion to occur at short interparticle separations if the dielectric constant of the medium is greater than the dielectric constants of both spheres, provide insights into the physics of charge separation in each sphere and of polarization in the medium behind such counterintuitive behaviour. 
\end{abstract}

\maketitle

\section{Introduction}

Electrostatic interactions are ubiquitous in nature, with examples spanning physical, chemical, and biological systems~\cite{messina2009electrostatics,PhyCheBiolevin2005strange,CheBiohonig1995classical,Biodoerr2006electrostatics,Biodoerr2017extending,Biokornyshev2007structure,ohshima2024fundamentals} and including binding processes in enzyme catalysis~\cite{warshel2006electrostatic}, solvent polarity of ionic liquids~\cite{exhallett2011room}, and the self-assembly of nanoparticles~\cite{exboles2016surface,barros2014dielectric}. At the atomistic or molecular level, electrostatic interactions are crucial in the bond-formation processes of ionic compounds, the attractive interactions of polar molecules, and the stabilization of large biomolecules such as proteins and DNAs~\cite{SPquiocho1987stabilization}. At the macroscopic level, electrostatic phenomena manifest themselves in a variety of scenarios, from triboelectricity to atmospheric lightning~\cite{Spguest1933static}. 

One of the fundamental topics of interest, relating to a variety of electrostatic phenomena in nature, concerns how a pair of polarizable particles at given electrostatic charges, dielectric constants and particle sizes and at a given interparticle separation interact. This is relevant to the formation of clouds and rains~\cite{Dropambaum2022enhanced}, aerosol growth in the atmosphere of Titan~\cite{Titanlindgren2017effect}, and climate change due to the lifting of mineral dust~\cite{Dustesposito2016role}. Surprisingly, for such a simple system, we are still far from a comprehensive understanding of the corresponding mechanisms of interparticle interactions, especially at short interparticle separations where non-linear effects are dominant due to multiple scattering of electric fields between the particles~\cite{freed2014perturbative}. 

While the classic Coulomb's law~\cite{jackson2021classical} describes the electrostatic force between two point charges in a vacuum, it does not necessarily describe the electrostatic force between two charged dielectric particles. In some cases, there could be a significant deviation from the Coulomb force at small separations, resulting in the occurrence of polarization-driven like-charge attraction~\cite{
FEMfeng2000electrostatic,MEbichoutskaia2010electrostatic,ICMxu2013electrostatic,freed2014perturbative,MSFqin2016theory,MSFqin2016image,EDlindgren2018electrostatic,MSFqin2019charge,PIchan2020theory,LCAli2024like,BEMgorman2024electrostatic}. On the other hand, charged particles immersed in an electrolyte could also exhibit anti-Coulomb behaviour, including cases of like-charge attraction ~\cite{LCAallahyarov1998attraction,LCAlevin1999nature,LCAlinse1999electrostatic,LCAmoreira2001binding,LCAnagornyak2009mechanism,LCAzhao2016like,LCAbuyukdagli2017like,LCAdos2019like,LCAwills2024anti,naturewang2024charge} and also cases of opposite-charge repulsion~\cite{OCRaranda1999electrostatic,OCRzhang2004phase,OCRleunissen2005ionic,LCAwills2024anti}, but such behaviour is not purely driven by the polarizability of the particles nor that of the surrounding medium. Such cases, where ionic effects are significant, could be investigated theoretically via mean-field approaches~\cite{Bbudkov2022modified,Bbukolesnikov2022electrosorption}, the variational-field theory~\cite{Bbudkov2023variational}, the statistical-field theory~\cite{Bburandyshev2023statistical}, the self-consistent field theory~\cite{Bbudkov2023macroscopic}, or a thermomechanical approach~\cite{Bbudkov2024surface}. While it is physically impossible for two dielectric particles in a vacuum to exhibit opposite-charge repulsion, by taking into account the polarizability of the medium it would be worthwhile to investigate whether opposite-charge repulsion could also occur as a pure consequence of polarization effects, as in the case of like-charge attraction between two dielectric particles in a vacuum. 

To understand the physical mechanisms behind the occurrence of polarization-driven like-charge attraction between two polarizable particles, much research has been conducted by means of 1) accurate evaluations of the interparticle force or energy \cite{MEbichoutskaia2010electrostatic,BCkhachatourian2014electrostatic,MElindgren2016progress,EDlindgren2018electrostatic,BClian2018polarization,LCAli2024like} that take into account all possible higher-order interactions, or 2) effective-dipole-based models\cite{PIchan2020theory} or computations\cite{EDlindgren2018electrostatic,LCAli2024like} that provide a physical picture of charge separation in each particle. For the former, a multiple scattering formalism~\cite{freed2014perturbative,MSFqin2016image,MSFqin2016theory,MSFqin2019charge} has been developed, where analytic solutions for generally many-body systems could be obtained through an application of the perturbative many-body expansion to the interaction energy. On the other hand, there have been various attempts to develop efficient numerical methods for accurate computations of force or energy: conventional approaches, such as the finite element method~\cite{FEMfeng2000electrostatic} and the boundary element method~\cite{BEMgorman2024electrostatic,BEMruan2022surface,BIEhassan2021integral,BIEbramas2021integral}, typically involve the discretization of a volume or surface. At small interparticle separations, such approaches become computationally expensive, because a high resolution of the discretization is required. For this reason, some more efficient approaches have been developed. These include 1) the image charge method ~\cite{ICMwang2013effects,ICMxu2013electrostatic} where, for each particle, the polarization field is approximated as a field generated by the image charges in the other particle and by the multiply-reflected image charges due to the presence of two interfaces, 2) the method of moments ~\cite{MEbichoutskaia2010electrostatic,MEderbenev2016electrostatic,MElindgren2016progress,MEsiryk2021charged} where the polarization-related potential is described in terms of a multipole expansion, the coefficients of which are determined by substituting the expansion into the Poisson equation and the corresponding boundary conditions, and 3) a hybrid computational method that employs the image charge method to compute the fast varying near-field interactions and the method of moments to solve the remaining smooth far-field interactions.


In this research, we used a combination of the 
hybrid computational method with the effective-dipole approach to investigate the possibility of polarization-induced opposite-charge repulsion between two dielectric spheres in close proximity inside a polarizable medium. This effective-dipole approach averages over not only the first-order dipole contributions but also all quadrupole and higher-order contributions. Our findings revealed that such a possibility exists if the dielectric constant of the medium is greater than the dielectric constants of both spheres, in which case the repulsive interaction corresponds to an anti-parallel alignment of the particles' effective dipoles \cite{EDlindgren2018electrostatic} and hence a dominant like-charge repulsion between the polarization charges of the particles. The problem of how the electrostatic interactions of charged dielectric particles can be mediated by the polarizability of the medium is of fundamental interest and importance to the scientific community. In this paper, we provide a unified understanding of how the counterintuitive polarization-driven phenomena of like-charge attraction and opposite-charge repulsion can be triggered, where our effective-dipole analysis provides insights into the physics of charge separation in each particle under the influence of a polarizable medium.

This paper is organized as follows. In Sec.~\ref{sec:model}, we present a theoretical model of two charged dielectric spheres, on which our research is based, along with a brief review of the hybrid computational method~\cite{HMgan2019efficient} that we employed to compute the interparticle force and the effective-dipole moment of each particle ~\cite{EDlindgren2018electrostatic}. In Sec.~\ref{sec:scenarios}, we present our findings for four different scenarios of interparticle interactions. In Sec.~\ref{sec:mechanims}, we discuss the mechanisms of interparticle interactions and the corresponding orientations and relative positions of effective dipoles for each scenario. In Sec.~\ref{sec:conclusion}, we summarize our findings and discuss their implications.

\section{Model and Methodology}~\label{sec:model}

\subsection{Model of dielectric spheres in a polarizable medium}

\begin{figure}
	\centering
	\includegraphics[scale=0.55]{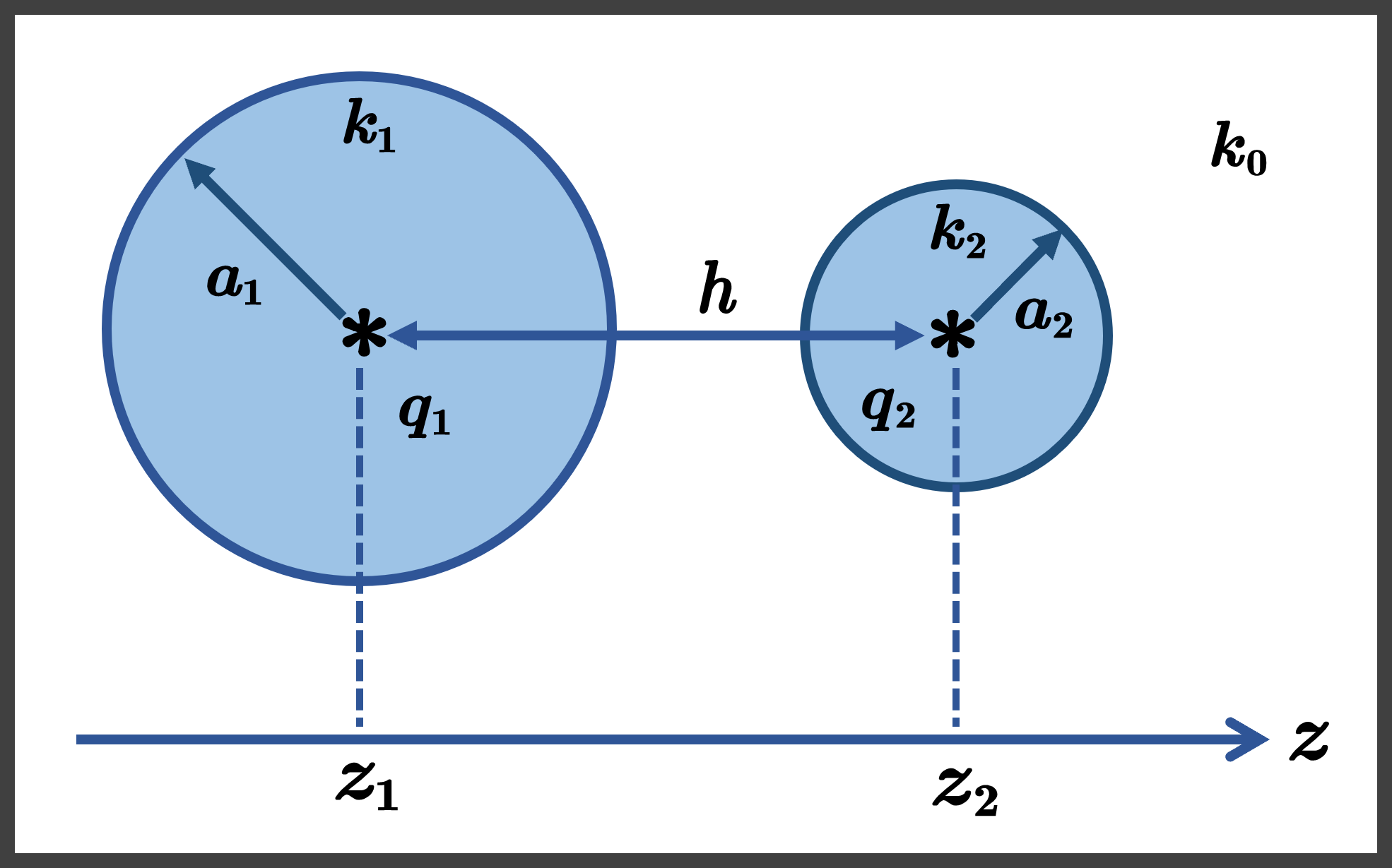}
	\caption{Schematic illustration of two dielectric spheres, immersed in a medium of dielectric constant $k_0$ and separated by a centre-to-centre separation $h$. For sphere $i\in\left\{1,2\right\}$ with a radius $a_i$ and a dielectric constant $k_i$, its free charge $q_i$ is concentrated at the particle's centre $\mathbf{z}_i$, and its bound charge spans the particle's surface.}		
	\label{fig1}
\end{figure}

Consider two dielectric spheres immersed in a polarizable medium (see Fig.\ref{fig1}) and separated by a centre-to-centre separation $h$. For sphere $i\in\left\{1,2\right\}$ with a radius $a_i$, its free charge $q_i$ is concentrated at the particle's centre $\mathbf{z}_i=\left[0, 0, z_i\right]$, and its bound charge spans the particle's surface. 
At any position $\mathbf{r}$, the volume density $\rho(\mathbf{r})$ of total charge is a sum of the volume density $\rho_{\text{f}}(\mathbf{r})$ of free charge and the volume density $\rho_{\text{b}}(\mathbf{r})$ of polarization-induced bound charge:
\begin{equation}
	\rho(\mathbf{r}) = \rho_{\text{f}}(\mathbf{r}) + \rho_{\text{b}}(\mathbf{r})\;,
\end{equation}
where
\begin{equation}
	\rho_{\text{f}}(\mathbf{r}) = \sum_{i=1}^{2} q_{i} \delta(\mathbf{r} - \mathbf{r}_{i})\;,
\end{equation}
The electrostatic energy $U$ of the system is given by\cite{HMgan2019efficient}
\begin{equation}\label{eq:energy}
	U = \frac{1}{2} \int \rho_{\text{f}}(\mathbf{r}) \phi(\mathbf{r}) d\mathbf{r}\;,
\end{equation}
where the electrostatic potential $\phi(\V r)$ is a sum of Coulomb contributions from all free charge and bound charge elsewhere in the system. It is worth pointing out that the electrostatic energy $U$ duly excludes the interaction between any given point charge and the potential it generates, where such self-interaction would inappropriately make the energy divergent. This potential approaches zero at any position far away from the pair of dielectric spheres. According to Gauss' law, we have
\begin{equation}\label{eq:alternative}
	-\grad^{2}\phi(\mathbf{r})=\frac{\rho(\mathbf{r})}{\epsilon_0}\;
\end{equation}
or equivalently
\begin{equation}
	-\grad \cdot\left[\epsilon_{\text{r}}(\mathbf{r}) \grad \phi(\mathbf{r})\right]=\frac{\rho_{\text{f}}(\mathbf{r})}{\epsilon_0}\;,
\end{equation}
where $\epsilon_{\text{r}}(\mathbf{r})$ is the relative permittivity and $\epsilon_0$ the permittivity of free space. Since the polarizable medium and the two dielectric spheres are each homogeneous, the position-dependent relative permittivity is given by 
\begin{equation}
	\epsilon_{\text{r}}(\mathbf{r})=\left\{
	\begin{array}{cl}
		k_{i}, &  \text{sphere $i$} \\
		\\
		k_0,  &  \text{medium} \\
	\end{array} \right.
\end{equation}
where $k_{i}$ is the dielectric constant of sphere $i$ and $k_0$ the dielectric constant of the medium. At the interfacial boundary between sphere $i$ and the medium, the interface condition 
\begin{equation}\label{eq:interfaceCond1}
    \begin{split}
        \phi(\mathbf{r}^-) &= \phi(\mathbf{r}^+)\;\\		
    \end{split}
\end{equation}
for a continuity of the electrostatic potential, and the interface condition
\begin{equation}\label{eq:interfaceCond2}
    \begin{split}
        k_{i} \frac{\partial \phi(\mathbf{r})}{\partial \V n}\big|_{\mathbf{r}=\mathbf{r}^-} &=\   k_0    \frac{\partial \phi(\mathbf{r})}{\partial \V{n}}\big|_{\mathbf{r}=\mathbf{r}^+}\;
    \end{split}
\end{equation}
for the absence of free surface charge, are satisfied. The superscripts $\mathbf{r}^+$ and $\mathbf{r}^-$ denote the limits of the position $\mathbf{r}$ as approached from, respectively, the medium and the sphere. This sphere-medium boundary is characterised by 
a surface density $\sigma_{\text{b}}(\mathbf{r})$ of bound charge, from which the effective-dipole moment of the sphere could be computed. With the above specification, our model is well-defined with the existence of a unique solution of $\phi(\mathbf{r})$ for any position $\mathbf{r}$. 

\subsection{Computational method}
Using spherical harmonics, the electrostatic potential inside the medium and that inside sphere $i$ can be expressed as
\begin{equation}\label{eq:hybridphi1}
    \begin{split}
        \phi_{0}(\mathbf{r}) &= \frac{1}{4\pi \epsilon_0 k_0}\sum_{i=1}^{2}\sum_{\ell,m}\frac{A_{i,\ell m}}{r_{i}^{\ell +1}}Y_{\ell}^{m}(\theta_{i},\psi_{i})\;\\		
    \end{split}
\end{equation}
and
\begin{equation}\label{eq:hybridphi2}
    \begin{split}
        \phi_{i}(\mathbf{r}) &= \frac{q_{i}}{4\pi \epsilon_{0}k_{i} |\mathbf{r}-\mathbf{z}_{i}|} + \frac{1}{4\pi\epsilon_0 k_0}\sum_{\ell,m}B_{i,\ell m} r_{i}^{\ell} Y_{\ell}^{m}(\theta_{i},\psi_{i})\;
    \end{split}
\end{equation}
respectively, where $Y_{\ell}^{m}(\theta_{i},\psi_{i})$ denotes the spherical harmonics function with degree $\ell$ and order $m$. The expansion coefficients and $A_{i,\ell m}$ and $B_{i,\ell m}$ were evaluated numerically via the aforementioned hybrid computational method ~\cite{HMgan2016hybrid, HMgan2019efficient}, and truncated up to some prescribed order $p$ to achieve a desired accuracy. Although we do not have general analytic solutions for the expansion coefficients $A_{i,\ell m}$ and $B_{i,\ell m}$, there is one special case where the coefficient can be determined analytically and is found to be $h$-independent ~\cite{HMgan2019efficient}: $A_{i,00}=q_i$. Any $h$-dependence of monopole strength arises from a screening effect of mobile ions, which is not considered in our study of purely polarization-driven interparticle interactions. Therefore the model considered in our study corresponds to the limiting case of zero ionic strength~\cite{FisherIonicMedium1994}. Upon a computation of the electrostatic energy $U$ using Eqs.~\eqref{eq:energy} and~\eqref{eq:hybridphi2}, the electrostatic force on sphere $i$ can be evaluated as follows:  
\begin{equation}\label{eq:forceFi}
\V F_i = -\grad_{\V z_i} U\;.
\end{equation}

For sphere $i$, the surface density of bound charge is given by
\begin{equation}\label{eq:sigmab}
	\sigma_{\text{b},i}(\mathbf{r}) = \frac{1}{4\pi\epsilon_0}\sum_{\ell,m} \left[(2 \ell +1)\frac{A_{i,\ell m}}{k_0 a_{i}^{\ell +2}}-\frac{q_{i}}{k_{i} a_{i}^2}\delta_{\ell 0}\right]Y_{\ell}^{m}(\theta_{i},\psi_{i})\;
\end{equation}
according to Eqs.~\eqref{eq:hybridphi1} and~\eqref{eq:hybridphi2} and the corresponding interfacial boundary condition. For the sphere's effective dipole~\cite{EDlindgren2018electrostatic}, the total positive polarization-charge
\begin{equation} \label{z_positive}
\widetilde{q}_{i}^{+}\coloneqq  \int\sigma_{\text{b},i}(\mathbf{r}) \mathds{1}_{\text{P}}(\sigma_{\text{b},i}(\mathbf{r})) dS_{i}\;\\
\end{equation}
and its charge-weighted position
\begin{equation} \label{r_positive}
\widetilde{\mathbf{r}}_{i}^{+} \coloneqq \frac{\int  \sigma_{\text{b},i}(\mathbf{r}) \mathds{1}_{\text{P}}(\sigma_{\text{b},i}(\mathbf{r})) \mathbf{r} dS_{i}}{\int\sigma_{\text{b},i}(\mathbf{r}) \mathds{1}_{\text{P}}(\sigma_{\text{b},i}(\mathbf{r})) dS_{i}}\;\\
\end{equation}
as well as the total negative polarization-charge 
\begin{equation} \label{z_negative}
\widetilde{q}_{i}^{-} \coloneqq \int\sigma_{\text{b},i}(\mathbf{r}) \mathds{1}_{\text{N}}(\sigma_{\text{b},i}(\mathbf{r})) dS_{i}\;\\
\end{equation}
and its charge-weighted position
\begin{equation} \label{r_negative}
\widetilde{\mathbf{r}}_{i}^{-} \coloneqq \frac{\int \sigma_{\text{b},i}(\mathbf{r})\mathds{1}_{\text{N}} (\sigma_{\text{b},i}(\mathbf{r}))\mathbf{r} dS_{i}}{\int\sigma_{\text{b},i}(\mathbf{r}) \mathds{1}_{\text{N}} (\sigma_{\text{b},i}(\mathbf{r})) dS_{i}}\;    
\end{equation}
can then be computed using Eq.~\eqref{eq:sigmab} as well as two indicator functions, defined as
\begin{equation}
	\mathds{1}_{\text{P}}(\sigma_{\text{b},i}(\mathbf{r}))\coloneqq \left\{
	\begin{array}{cl}
		1, &  \sigma_{\text{b},i}(\mathbf{r})>0 \\
		\\
		0,  &  \sigma_{\text{b},i}(\mathbf{r})\leq 0\\
	\end{array} \right.
\end{equation}
and
\begin{equation}
	\mathds{1}_{\text{N}}(\sigma_{\text{b},i}(\mathbf{r}))  \coloneqq \left\{
	\begin{array}{cl}
		1, &  \sigma_{\text{b},i}(\mathbf{r})<0 \\
		\\
		0,  &  \sigma_{\text{b},i}(\mathbf{r})\geq 0\\
	\end{array} \right.
\end{equation}
respectively. Any polarization effect, which by definition corresponds to a spatial separation of positive and negative charge, does not contribute to the net charge of a sphere, and therefore the total polarization charge of a sphere must be zero, i.e. 
\begin{equation} \label{chargeneutral}
\widetilde{q}_{i}^{+}+\widetilde{q}_{i}^{-}=0\;\\
\end{equation}
Eq. \eqref{chargeneutral} has been verified in all of our numerical simulations. For sphere $i$, the absolute quantity $\widetilde{q}_{i} \coloneqq |\widetilde{q}_{i}^{\pm}|$ is the magnitude of induced charge, and the positions $\widetilde{\mathbf{r}}_{i}^{\pm}$ are the charge-weighted average positions of the two types of induced charge. Due to azimuthal symmetry, the induced charges are all located on the $z$-axis such that the positions $\widetilde{\mathbf{r}}_{i}^{\pm}$ are given by $\widetilde{\mathbf{r}}_{i}^{+} \coloneqq [0,0,\widetilde{z}_{i}^{+}]$ and $\widetilde{\mathbf{r}}_{i}^{-} \coloneqq [0,0,\widetilde{z}_{i}^{-}]$, respectively.

\section{Scenarios of interparticle interactions} ~\label{sec:scenarios}

\begin{figure}
	\centering
	\includegraphics[scale=0.79]{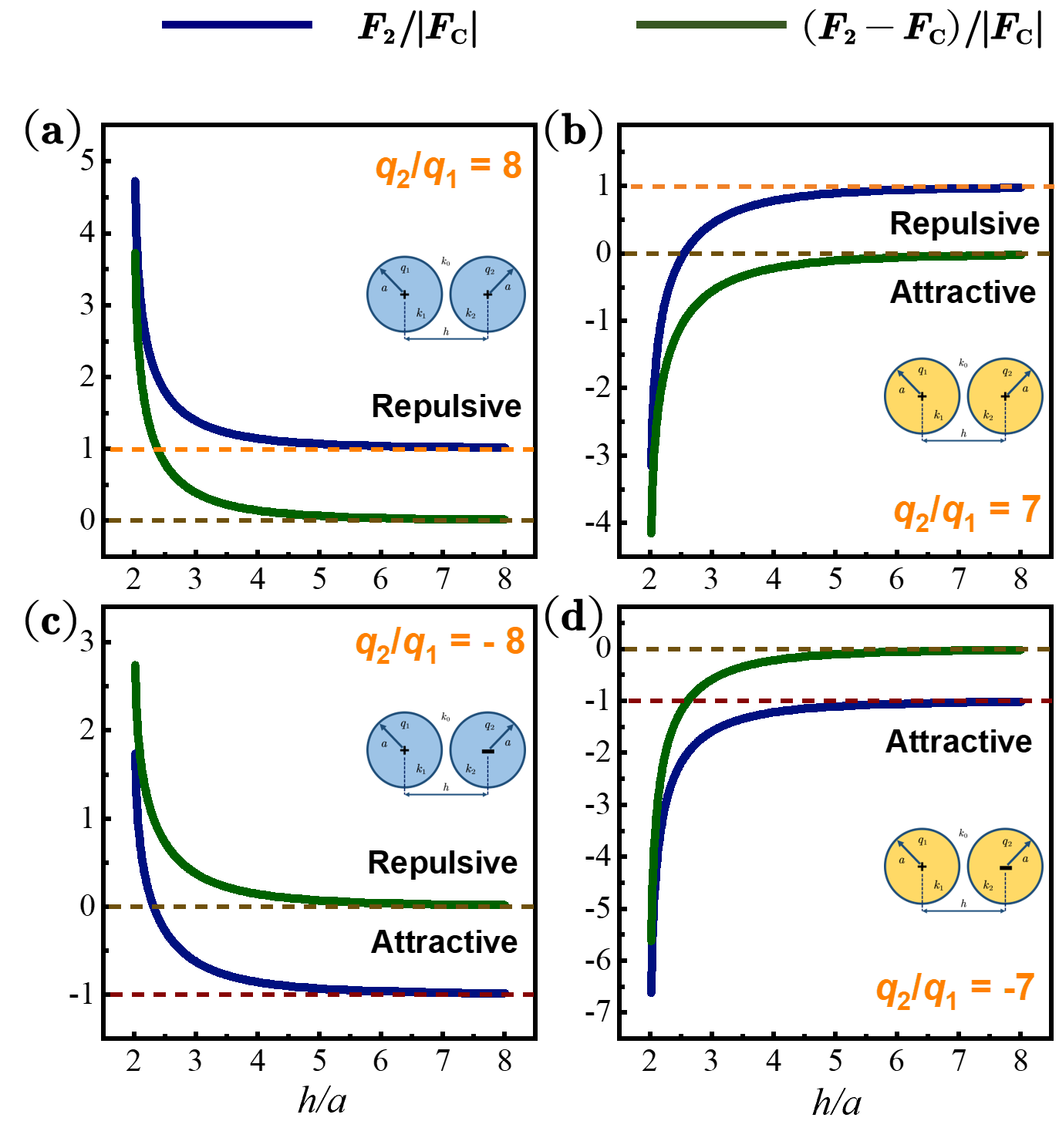}
	\caption{Rescaled electrostatic force and rescaled polarization force on sphere 2 for a pair of equal-sized dielectric spheres inside a polarizable medium: a) $k_1/k_0=0.001$, $k_2/k_0=0.08$ and $q_2/q_1=8$ (scenario A); (b) $k_1/k_0=30$, $k_2/k_0=4$ and $q_2/q_1=7$ (scenario B); (c) $k_1/k_0=0.001$, $k_2/k_0=0.08$ and $q_2/q_1=-8$ (scenario C); (d) $k_1/k_0=30$, $k_2/k_0=4$ and $q_2/q_1=-7$ (scenario D).}		
	\label{equalsize}
\end{figure}

\begin{figure}
	\centering
	\includegraphics[scale=0.79]{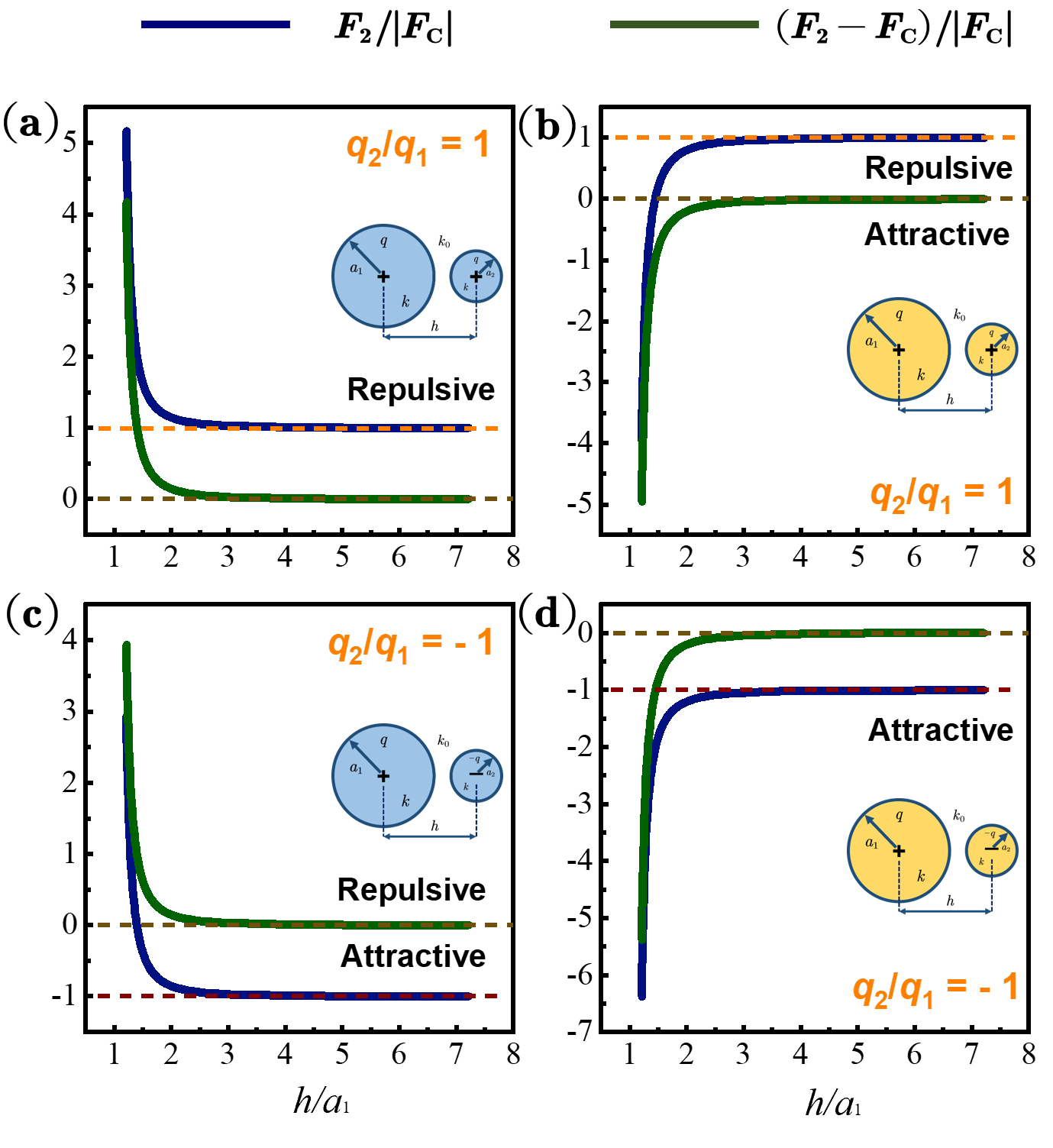}
	\caption{Rescaled electrostatic force and rescaled polarization force on sphere 2 for a pair of unequal-sized dielectric spheres at $|q_1|=|q_2|$, $a_2/a_1=0.2$ and $k_1/k_0=k_2/k_0=k$ inside a polarizable medium: (a) $q_1=q_2$ and $k=0.25$ (scenario A); (b) $q_1=q_2$ and $k=4$ (scenario B); (c) $q_1=-q_2$ and $k=0.25$ (scenario C); (d) $q_1=-q_2$ and $k=4$ (scenario D).}
	\label{unequalsize}
\end{figure}

\begin{figure}
    \centering
    \includegraphics[scale=0.5]{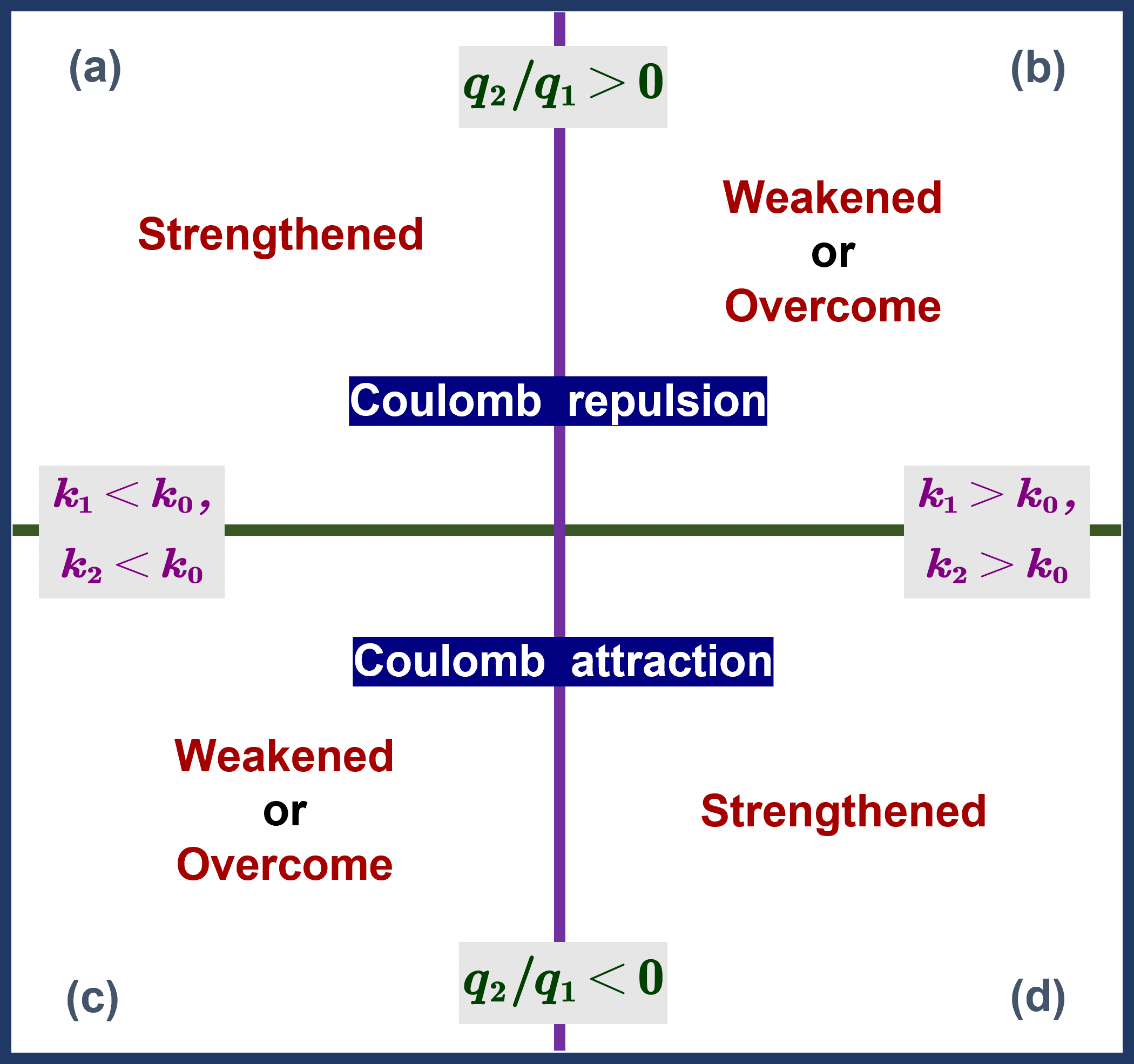}
    \caption{Four scenarios of electrostatic interactions under consideration: (a) $k_1<k_0$, $k_2<k_0$ and $q_2/q_1>0$ (scenario A); (b) $k_1>k_0$, $k_2>k_0$ and $q_2/q_1>0$ (scenario B); (c) $k_1<k_0$, $k_2<k_0$ and $q_2/q_1<0$ (scenario C); (d) $k_1>k_0$, $k_2>k_0$ and $q_2/q_1<0$ (scenario D).}
    \label{figclassification}
\end{figure}

To investigate how the electrostatic interaction between two charged dielectric spheres depends on the dielectric constants $k_i$, free charges $q_i$ and radii $a_i$ of the spheres ($i=1,2$) and on the dielectric constant $k_0$ of the polarizable medium, we first consider two types of scenarios: (i) two equal-sized spheres with different dielectric constants $k_i$ and charge magnitudes $|q_i|$, and (ii) two unequal-sized spheres sharing the same dielectric constant $k_i$ and charge magnitude $|q_i|$. For each type of scenarios, we consider how the total electrostatic force and polarization force vary with the interparticle separation at four possible combinations of charge ratio ($q_2/q_1>0$ for like-charge interactions, or $q_2/q_1<0$ for opposite-charge interactions) and medium polarizability ($k_i<k_0$ for a more polarizable medium, or $k_i>k_0$ for a less polarizable medium): (a) $q_2/q_1>0$ and $k_i<k_0$ (scenario A); (b) $q_2/q_1>0$ and $k_i>k_0$ (scenario B); (c) $q_2/q_1<0$ and $k_i<k_0$ (scenario C); (d) $q_2/q_1<0$ and $k_i>k_0$ (scenario D). Without loss of generality, we adopted the parameter values $k_0=1$, $q_1=1\text{e}$ and $a_1=1\mu\text{m}$ in our numerical simulations.

Fig. \ref{equalsize} shows how the rescaled electrostatic force and rescaled polarization force vary with the interparticle separation for cases of equal-sized spheres with different dielectric constants $k_i$ and charge magnitudes $|q_i|$. For scenarios A and C, where the dielectric constants of the spheres are smaller than that of the polarizable medium, the polarization force is all the way repulsive, regardless of whether the spheres exhibit like-charge or opposite-charge interactions. For scenario A, the Coulomb force is inherently repulsive, and the polarization force simply enhances the repulsive nature of the interparticle interaction. For scenario C, the Coulomb force is inherently attractive, and the polarization force results in a switch of the total electrostatic force from attraction to repulsion for decreasing separation. For scenarios B and D, where the dielectric constants of the spheres are larger than that of the polarizable medium, the polarization force is all the way attractive. For scenario B, the Coulomb force is inherently repulsive, and the polarization force results in a switch of the total electrostatic force from repulsion to attraction for decreasing separation. For scenario D, the Coulomb force is inherently attractive, and the polarization force simply enhances the attractive nature of the interparticle interaction. For all cases, at large interparticle separations, the polarization force approaches zero and the total electrostatic force becomes practically equal to the Coulomb force. Fig. \ref{unequalsize} shows corresponding results for cases of unequal-sized spheres sharing the same dielectric constant $k_i$ and charge magnitude $|q_i|$. The results, which display the same qualitative features as those shown in Fig. \ref{equalsize}, suggest a universal picture of electrostatic interactions upon the dominant polarization of one sphere by the other, as summarized in Fig. \ref{figclassification}:

If the dielectric constant of the surrounding medium exceeds the dielectric constants of both spheres, i.e. $k_1<k_0$ and $k_2<k_0$, the polarization within the system results in an additional repulsive interaction that either strengthens the Coulomb repulsion between two like-charged particles ($q_2 / q_1 >0$) in scenario A or weakens (or even overcomes) the Coulomb attraction between two oppositely charged particles ($q_2 / q_1 <0$) in scenario C. Else if the dielectric constant of the surrounding medium is smaller than the dielectric constants of both spheres, i.e. $k_1>k_0$ and $k_2>k_0$, the polarization within the system results in an additional attractive interaction that either weakens (or even overcomes) the Coulomb repulsion between two like-charged particles ($q_2 / q_1 >0$) in scenario B or strengthens the Coulomb attraction between two oppositely-charged particles ($q_2 / q_1 <0$) in scenario D.

\section{Mechanisms of interparticle interactions}\label{sec:mechanims}

Without loss of generality, we have conducted an effective-dipole analysis of the mechanisms of interparticle interactions for cases of two equal-sized spheres with different dielectric constants $k_i$ and charge magnitudes $|q_i|$, with a focus on how the nature of interaction varies as the spheres approach each other. For each scenario, we consider various combinations of $q_i$ and $k_i$ that correspond to a dominant mechanism of polarization of sphere 1 on the left by sphere 2 on the right (Fig. \ref{fig1}), and look into how our results vary with the strength of such polarization for decreasing interparticle separation.

~\\
    \textbf{Cases of a less polarizable medium}
~\\

If two charged spheres are placed in a less polarizable medium, for example in a vacuum, the effective dipole in the strongly polarized sphere 1 must align along the direction of the strong Coulomb field from sphere 2. If sphere 2 is positively charged, as in scenario B described below for two positively charged spheres, the induced negative charge in sphere 1 would be located closer to sphere 2 than its positive counterpart in the same sphere. Conversely, if sphere 2 is negatively charged, as in scenario D described below for a positively charged sphere 1 and a negatively charged sphere 2, the induced positive charge in sphere 1 would be located closer to sphere 2 than its negative counterpart in the same sphere. Scenarios B and D could be realized experimentally in a vacuum ($k_0=1$) with a lanthana-based sphere 1 ($k_1 \sim 30$) and a silica-based sphere 2 ($k_2 \sim 4$).

~\\
\textbf{Scenario B:  Like-charge repulsion or attraction in a less polarizable medium}
~\\

\begin{figure*}
	\centering
	\includegraphics[]{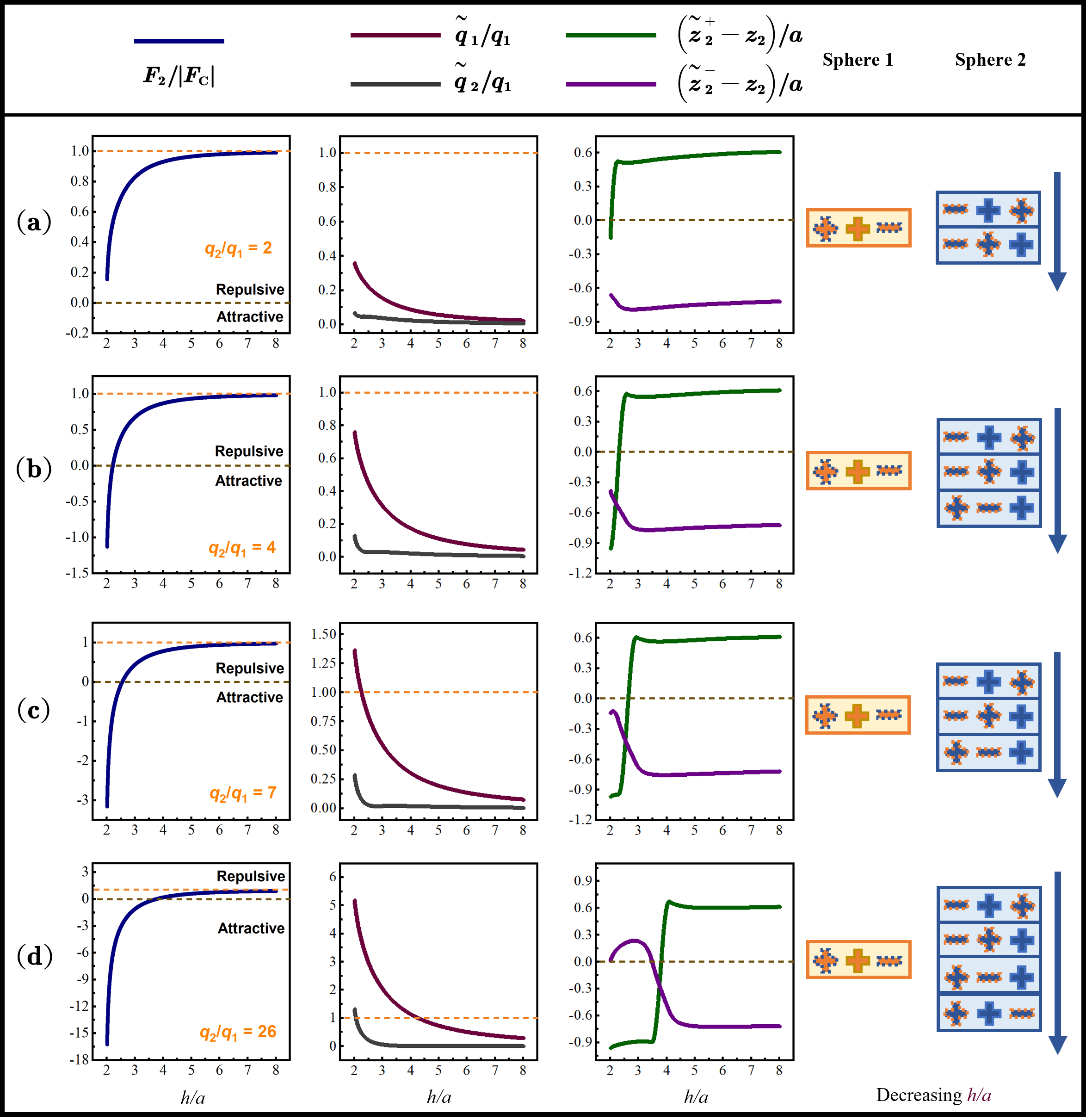}
	\caption{\textbf{Scenario B}: Like-charge repulsion or attraction for spheres immersed in a less polarizable medium at $k_1/k_0=30$ and $k_2/k_0=4$. The rows from top to bottom document results at respectively four different charge ratios: (a) $q_2/q_1=2$, (b) $q_2/q_1=4$, (c) $q_2/q_1=7$, and (d) $q_2/q_1=26$. The columns from left to right document results for (1) $F_2/\abs{F_\text{C}}$ ($F_2$ is the electrostatic force on sphere 2 and $F_\text{C}$ is the Coulomb force), (2) $\widetilde{q}_i/q_1$ (relative magnitude of induced charge in each sphere with respect to $q_1$), (3) $(\widetilde{z}_2^{\pm}-z_2)/a$ (relative positions of positive and negative induced charge in sphere 2 with respect to the sphere's centre) and (4) a pictorial representation of positional order of induced charge and free charge in each sphere. In columns 1 to 3, the horizontal axis is set as $h/a$, representing a relative centre-to-centre separation with respect to the radius of the equal-sized spheres. In column 4, any induced charge is represented by a symbol with a dashed boundary, and any free charge is represented by a symbol with a solid boundary.}	
	\label{fig2}
\end{figure*}

\begin{figure*}
    \centering
    \includegraphics[]{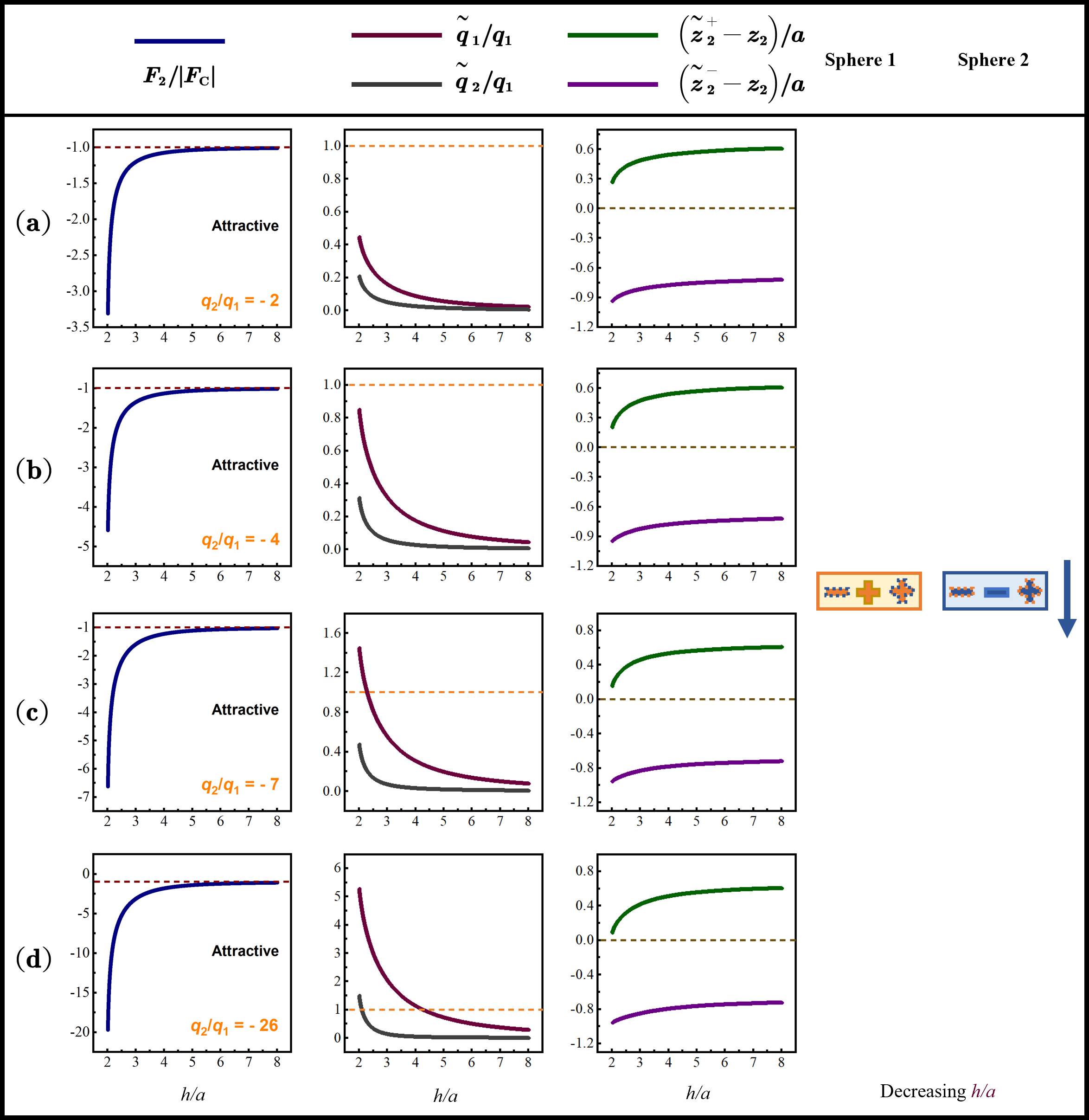}
    \caption{\textbf{Scenario D}: Opposite-charge attraction for spheres immersed in a less polarizable medium at $k_1/k_0=30$ and $k_2/k_0=4$. The rows from top to bottom document results at respectively four different charge ratios: (a) $q_2/q_1=-2$, (b) $q_2/q_1=-4$, (c) $q_2/q_1=-7$, and (d) $q_2/q_1=-26$. The columns from left to right document results for (1) $F_2/\abs{F_\text{C}}$ ($F_2$ is the electrostatic force on sphere 2 and $F_\text{C}$ is the Coulomb force), (2) $\widetilde{q}_i/q_1$ (relative magnitude of induced charge in each sphere with respect to $q_1$), (3) $(\widetilde{z}_2^{\pm}-z_2)/a$ (relative positions of positive and negative induced charge in sphere 2 with respect to the sphere's centre) and (4) a pictorial representation of positional order of induced charge and free charge in each sphere. In columns 1 to 3, the horizontal axis is set as $h/a$, representing a relative centre-to-centre separation with respect to the radius of the equal-sized spheres. In column 4, any induced charge is represented by a symbol with a dashed boundary, and any free charge is represented by a symbol with a solid boundary.
}	
	\label{klargen}
\end{figure*}

\begin{figure}
    \centering
    \includegraphics[scale=0.87]{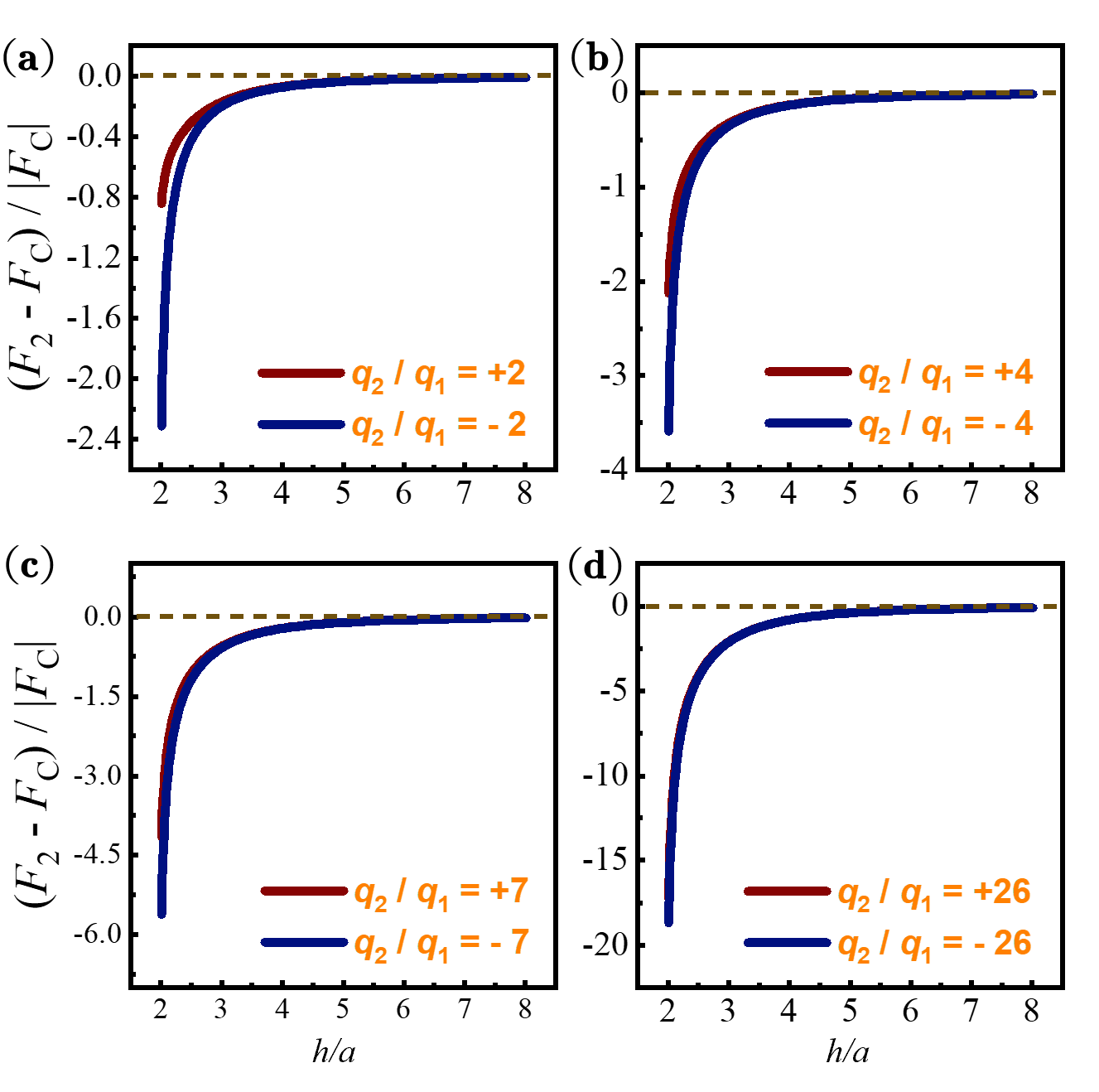}
    \caption{Plot of the rescaled polarization force $(F_2-F_{\mathrm{C}})/\abs{F_{\mathrm{C}}}$ against the rescaled interparticle separation $h/a$ for cases of like-charge interactions as well as cases of opposite-charge interactions, at $k_1/k_0=30$ and $k_2/k_0=4$. The sub-figures (a) to (d) document results at $q_2/q_1=\pm 2$, $q_2/q_1=\pm 4$, $q_2/q_1=\pm 7$ and $q_2/q_1=\pm 26$, respectively. In all cases, the polarization force is attractive, be it for cases of like-charge interactions or opposite-charge interactions.}
    \label{klargec}
\end{figure}

For scenario B we consider like-charge interactions for respectively the charge ratios (a) $q_2/q_1 = 2$, (b) $q_2/q_1 =4$, (c) $q_2/q_1 =7$, and (d) $q_2/q_1 =26$ with $q_1=1\text{e}$, at the dielectric-constant ratios $k_1/k_0 = 30$ and $k_2/k_0 = 4$ with $k_0 =1$. The results are summarized in Fig.~\ref{fig2}. As shown in the first column, for decreasing interparticle separation, there is generally an increasing deviation of the electrostatic force $F_2$ on sphere 2 from the Coulomb force $F_\text{C}$, due to a strengthening of polarization-induced attraction. At sufficiently large absolute values of $q_2/q_1$ (e.g. $q_2/q_1 =4$, $q_2/q_1 =7$ and $q_2/q_1 =26$), the Coulomb repulsion is overtaken by the polarization-induced attractive force at short interparticle separations such that the overall interparticle interaction becomes counterintuitively attractive. 

As shown in the second column, for decreasing interparticle separation, the magnitude of induced charge in each sphere as relative to $q_{1}$, i.e. $\widetilde{q}_{i}/q_{1}$, generally increases. For the four different values of $q_2/q_1$, we observed qualitative differences in the variation of $\widetilde{q}_{i}/q_{1}$ for decreasing interparticle separation: (a) At $q_2/q_1 = 2$, neither the value of $\widetilde{q}_{1}/q_{1}$ nor that of $\widetilde{q}_{2}/q_{1}$ exceeds unity, and the interparticle interaction remains repulsive all the way; (b) At $q_2/q_1 =4$, the magnitudes of induced charge in both spheres are still all the way smaller than $q_1$, yet the value of $\widetilde{q}_{1}/q_{1}$ is not far from unity at short interparticle separations. This corresponds to a sufficiently strong effect of polarization for the occurrence of like-charge attraction; (c) At $q_2/q_1 =7$, the value of $\widetilde{q}_{1}/q_{1}$ exceeds unity at small interparticle separations but the value of $\widetilde{q}_{2}/q_{1}$ is still all the way smaller than unity, and like-charge attraction occurs at short interparticle separations; (d) At $q_2/q_1 =26$, both the value of $\widetilde{q}_{1}/q_{1}$ and that of $\widetilde{q}_{2}/q_{1}$ exceed unity at certain interparticle separations, and like-charge attraction occurs at short interparticle separations.

As shown in the third column, the positions of induced positive charge and induced negative charge in sphere 2 as relative to the sphere’s centre exhibit a non-trivial variation for decreasing interparticle separation, at all four values of $q_2/q_1$. The results are translated into a pictorial representation in the fourth column, which illustrates the positional order of induced charge and free charge in each sphere. While the positional order of charge in sphere 1 is the same for any interparticle separation or charge ratio, i.e. with the induced positive charge on the left, the free positive charge in the middle, and the induced negative charge on the right, the positional order of charge in sphere 2 exhibits an interesting dependence on the interparticle separation and on the charge ratio, as discussed below:

At (a) $q_2/q_1=2$, the dominant polarization of sphere 1 by sphere 2 is relatively weak at large interparticle separations, such that the electrostatic force $F_2$ on sphere 2 is almost identical to the Coulomb force $F_\text{C}$. For decreasing interparticle separation, the mild enhancement in the induced negative charge $\widetilde{q}_{1}^{-}$ in sphere 1 results in an attraction of the induced positive charge $\widetilde{q}_{2}^{+}$ in sphere 2 from the rightmost position to the middle position in the corresponding pictorial representation. Despite such an alteration in charge distribution, the overall interparticle interaction is all the way repulsive, because the repulsive interaction between the induced negative charge $\widetilde{q}_{1}^{-}$ in sphere 1 and the induced negative charge $\widetilde{q}_{2}^{-}$ in sphere 2 remains dominant. 

At (b) $q_2/q_1=4$ and (c) $q_2/q_1=7$, the dominant polarization of sphere 1 by sphere 2 is generally stronger than the case of $q_2/q_1=2$. For decreasing interparticle separation, the significant enhancement in the induced negative charge $\widetilde{q}_{1}^{-}$ in sphere 1 results in an attraction of the induced positive charge $\widetilde{q}_{2}^{+}$ in sphere 2 from the rightmost position to the leftmost position in the corresponding pictorial representation. This results in a reversal of polarization in sphere 2 and a switch of the electrostatic force from repulsive to attractive, in which case the attractive interaction between the induced negative charge $\widetilde{q}_{1}^{-}$ in sphere 1 and the induced positive charge $\widetilde{q}_{2}^{+}$ in sphere 2 becomes dominant.

At (d) $q_2/q_1 =26$, the dominant polarization of sphere 1 by sphere 2 is stronger than the three cases described above. For decreasing interparticle separation, the significant enhancement in the induced negative charge $\widetilde{q}_{1}^{-}$ in sphere 1 not only results in an attraction of the induced positive charge $\widetilde{q}_{2}^{+}$ in sphere 2 from the rightmost position to the leftmost position in the corresponding pictorial representation, but also a repulsion of the induced negative charge $\widetilde{q}_{2}^{-}$ in sphere 2 from the leftmost position to the rightmost position. Like the cases of $q_2/q_1=4$ and $q_2/q_1=7$, this results in a reversal of polarization in sphere 2 and a switch of the electrostatic force from repulsive to attractive, in which case the attractive interaction between the induced negative charge $\widetilde{q}_{1}^{-}$ in sphere 1 and the induced positive charge $\widetilde{q}_{2}^{+}$ in sphere 2 becomes dominant.

~\\
\textbf{Scenario D: Opposite-charge attraction in a less polarizable medium}
~\\

To investigate how two oppositely-charged particles in a less polarizable medium interact and facilitate a comparison with scenario B, we have repeated the numerical investigations described above with a sign inversion of $q_2$ and with all other parameters remaining unchanged. The charge ratios considered are therefore (a) $q_2/q_1=-2$, (b) $q_2/q_1=-4$, (c) $q_2/q_1=-7$, and (d) $q_2/q_1=-26$, respectively, with $q_1=1\text{e}$. The results are summarized in Fig. \ref{klargen}. As shown in the first column, the electrostatic force between the two spheres is uniformly attractive, regardless of the charge ratio nor the interparticle separation. Such attraction, which diverges from the conventional attraction between opposite charges as described by Coulomb's law, is generally an enhancement of the Coulomb attraction. The larger the absolute value of the charge ratio, the stronger the enhancement. For decreasing interparticle separation, the electrostatic force $F_2$ on sphere 2 generally becomes more attractive and exhibits a greater deviation from the Coulomb force $F_\text{C}$, due to a strengthening of polarization-induced attraction. The fact that both the Coulomb force and the polarization-induced interaction are attractive in nature, as illustrated in Fig.~\ref{klargec} for the latter, excludes any possibility of opposite-charge repulsion.

Regardless of the charge ratio nor the interparticle separation, the positional order of charge in either sphere is fixed, i.e. with the induced negative charge on the left, the free positive or negative charge in the middle, and the induced positive charge on the right. This is because, for decreasing interparticle separation at any given charge ratio, the enhancement in the induced positive charge $\widetilde{q}_{1}^{+}$ in sphere 1 results in a further stabilization of the leftmost position of the induced negative charge $\widetilde{q}_{2}^{-}$ and the rightmost position of the induced positive charge $\widetilde{q}_{2}^{+}$ in the pictorial representation of sphere 2.

In general, if the dielectric constant of the surrounding medium is smaller than the dielectric constants of two like-charged spheres, i.e. $k_1>k_0$ and $k_2>k_0$, the force $F_2$ acting on sphere 2 is a mitigation of the Coulomb repulsion by a polarization-induced attractive force. Else if the spheres are oppositely charged, the force $F_2$ experienced by sphere 2 is an enhancement of the Coulomb attraction by a polarization-induced attractive force. It is worth noting that a sign inversion of $q_2$ has little impact on the polarization-induced attractive force, as illustrated in Fig.~\ref{klargec} for the charge ratios $q_2/q_1=\pm 2$, $q_2/q_1=\pm 4$, $q_2/q_1=\pm 7$ and $q_2/q_1=\pm 26$.

~\\
\textbf{Cases of a more polarizable medium}
~\\

A possible experimental realization is the placement of a non-polarizable sphere 1 ($k_1 \sim 1$) and a water-based sphere 2 ($k_2 \sim 80$), inside a superdielectric medium~\cite{Superfromille2014super} of dielectric constant $k_0=1000$. In such a strongly polarizable medium, the region surrounding the strongly polarizing sphere 2 is filled with induced charge. As a result of significant influence from such induced charge, the effective dipole in sphere 1 must align in a direction opposite to the strong Coulomb field from sphere 2.

~\\
\textbf{Scenario A: Like-charge repulsion in a more polarizable medium}
~\\

\begin{figure*}
    \centering
    \includegraphics[]{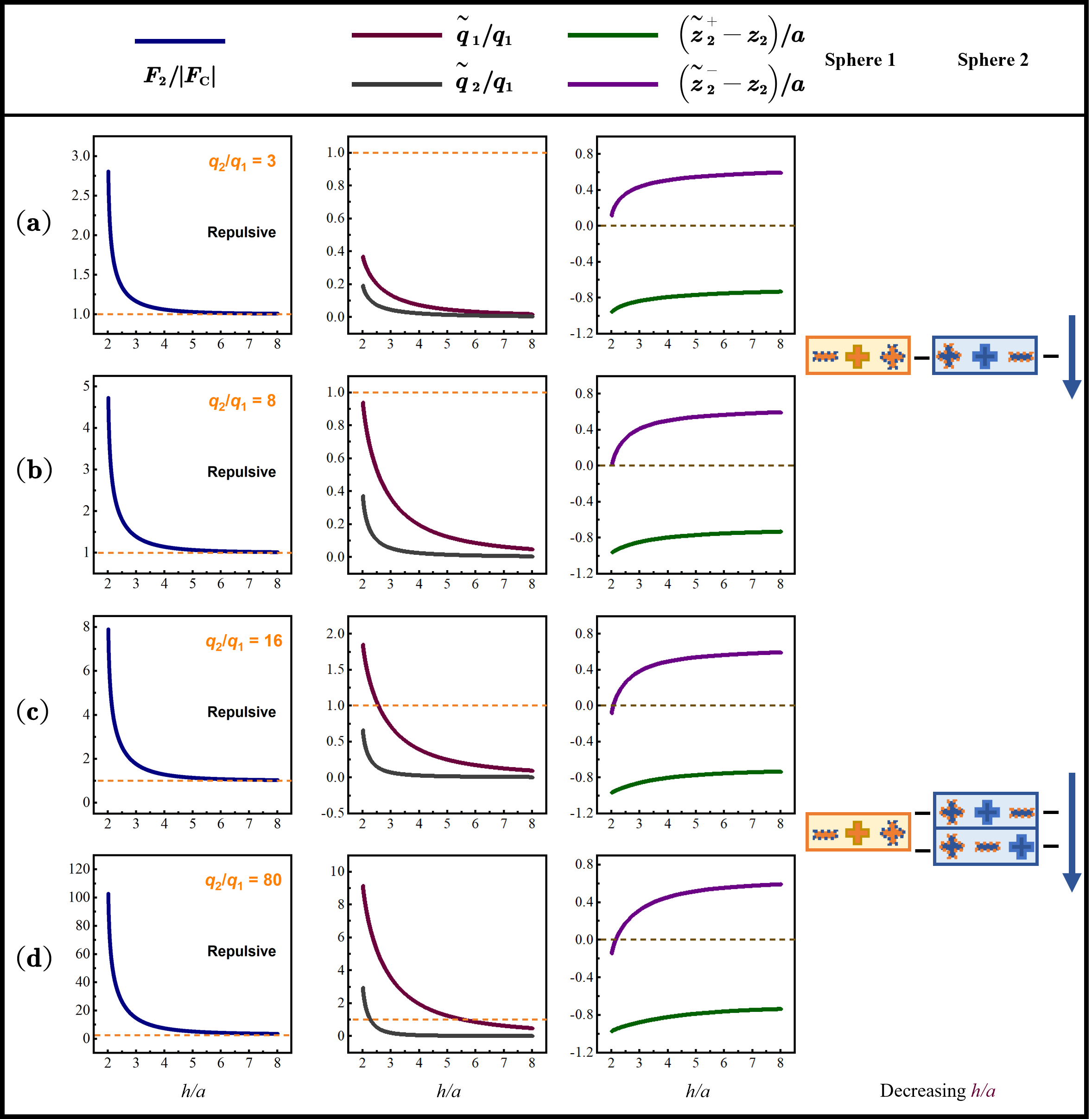}
    \caption{\textbf{Scenario A}: Like-charge repulsion for spheres immersed in a more polarizable medium at $k_1/k_0=0.001$ and $k_2/k_0=0.08$. The rows from top to bottom document results at respectively four different charge ratios: (a) $q_2/q_1=3$, (b) $q_2/q_1=8$, (c) $q_2/q_1=16$, and (d) $q_2/q_1=80$. The columns from left to right document results for (1) $F_2/\abs{F_\text{C}}$ ($F_2$ is the electrostatic force on sphere 2 and $F_\text{C}$ is the Coulomb force), (2) $\widetilde{q}_i/q_1$ (relative magnitude of induced charge in each sphere with respect to $q_1$), (3) $(\widetilde{z}_2^{\pm}-z_2)/a$ (relative positions of positive and negative induced charge in sphere 2 with respect to the sphere's centre) and (4) a pictorial representation of positional order of induced charge and free charge in each sphere. In columns 1 to 3, the horizontal axis is set as $h/a$, representing a relative centre-to-centre separation with respect to the radius of the equal-sized spheres. In column 4, any induced charge is represented by a symbol with a dashed boundary, and any free charge is represented by a symbol with a solid boundary.}	
	\label{ksmallp}
\end{figure*}

\begin{figure*}
	\centering
	\includegraphics[]{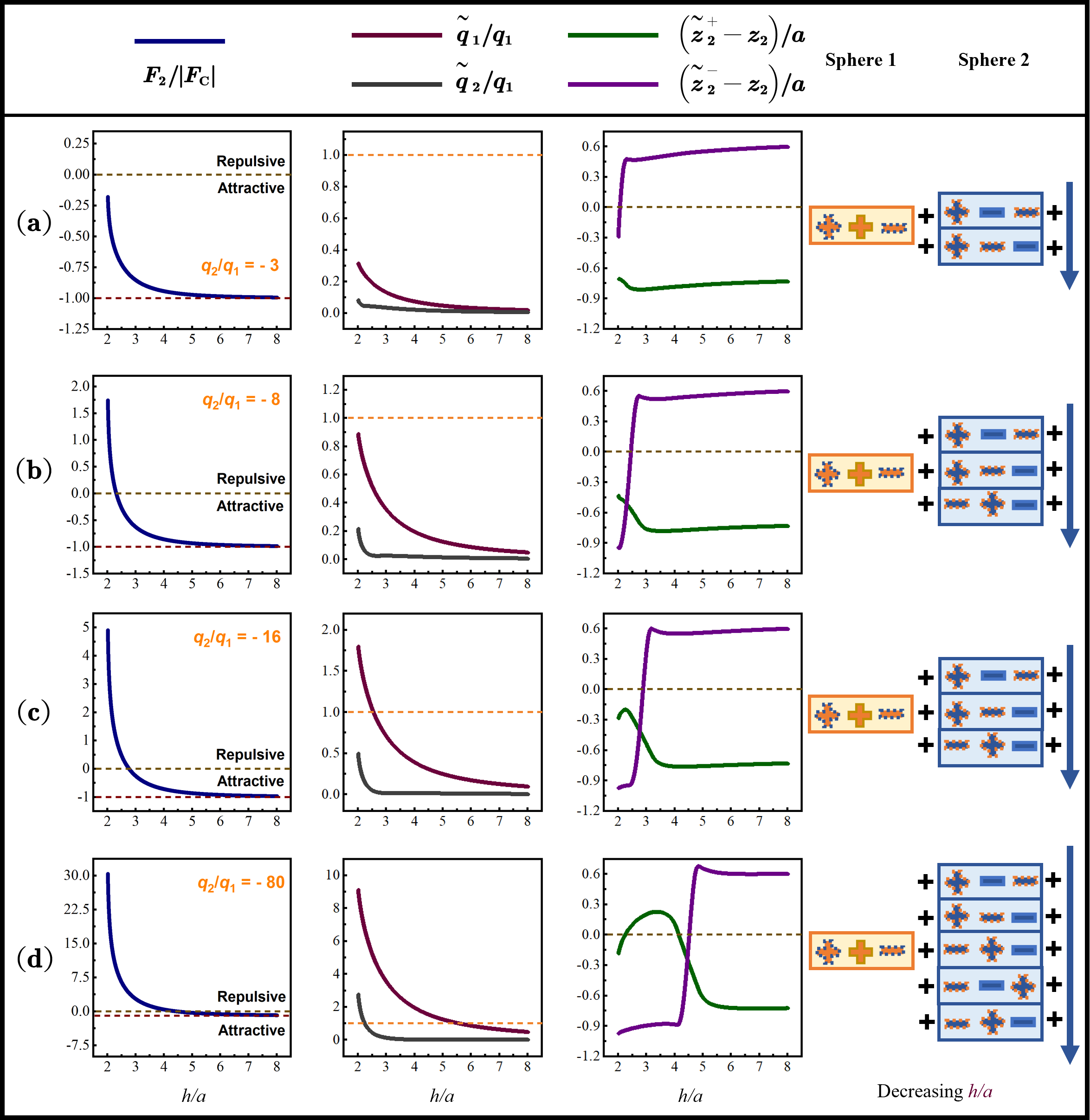}
	\caption{\textbf{Scenario C}: Opposite-charge attraction or repulsion for spheres immersed in a more polarizable medium at $k_1/k_0=0.001$ and $k_2/k_0=0.08$. The rows from top to bottom document results at respectively four different charge ratios: (a) $q_2/q_1=-3$, (b) $q_2/q_1=-8$, (c) $q_2/q_1=-16$, and (d) $q_2/q_1=-80$. The columns from left to right document results for (1) $F_2/\abs{F_\text{C}}$ ($F_2$ is the electrostatic force on sphere 2 and $F_\text{C}$ is the Coulomb force), (2) $\widetilde{q}_i/q_1$ (relative magnitude of induced charge in each sphere with respect to $q_1$), (3) $(\widetilde{z}_2^{\pm}-z_2)/a$ (relative positions of positive and negative induced charge in sphere 2 with respect to the sphere's centre) and (4) a pictorial representation of positional order of induced charge and free charge in each sphere. In columns 1 to 3, the horizontal axis is set as $h/a$, representing a relative centre-to-centre separation with respect to the radius of the equal-sized spheres. In column 4, any induced charge is represented by a symbol with a dashed boundary, and any free charge is represented by a symbol with a solid boundary.}		
	\label{ksmalln}
\end{figure*}

\begin{figure}
	\centering
	\includegraphics[scale=1.1]{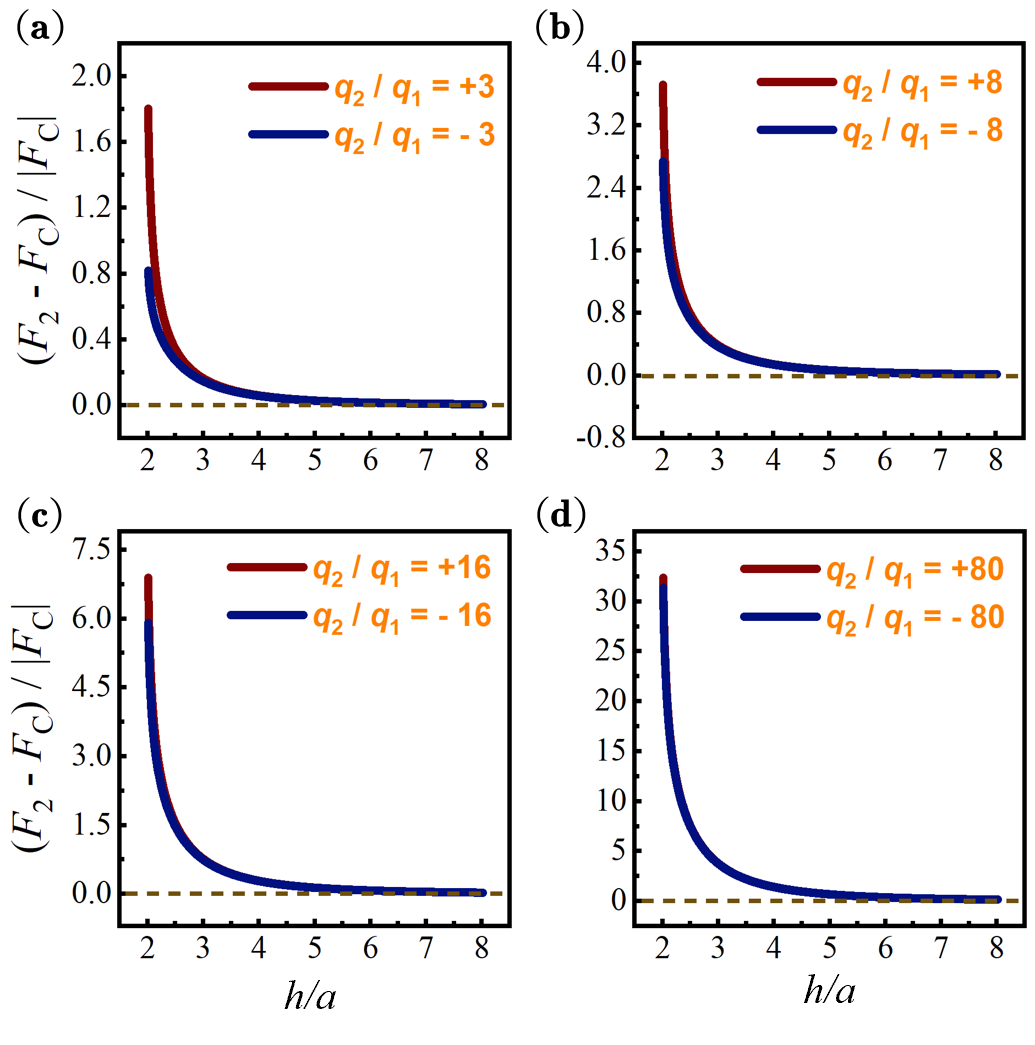}
	\caption{Plot of the rescaled polarization force $(F_2-F_{\mathrm{C}})/\abs{F_{\mathrm{C}}}$ against the rescaled interparticle separation $h/a$ for cases of like-charge interactions as well as cases of opposite-charge interactions, at $k_1/k_0=0.001$ and $k_2/k_0=0.08$. The sub-figures (a) to (d) document results at $q_2/q_1=\pm 3$, $q_2/q_1=\pm 8$, $q_2/q_1=\pm 16$ and $q_2/q_1=\pm 80$, respectively. In all cases, the polarization force is repulsive, be it for cases of like-charge interactions or opposite-charge interactions.}		
	\label{ksmallc}
\end{figure}

We have conducted numerical investigations for four different charge ratios at (a) $q_2/q_1=3$, (b) $q_2/q_1=8$, (c) $q_2/q_1=16$, and (d) $q_2/q_1=80$, respectively, with $q_1=1\text{e}$. The results, which are summarized in Fig. \ref{ksmallp}, are analogous to scenario D, where in either scenario the corresponding Coulomb interaction is enhanced. As an enhancement of the Coulomb repulsion, the electrostatic force between the two spheres is uniformly repulsive, where the same repulsive nature of the Coulomb force and the polarization-induced interaction (Fig.~\ref{ksmallc}) excludes any possibility of like-charge attraction.

In this scenario, the positional order of charge in sphere 1 is fixed, i.e. with the induced negative charge on the left, the free positive charge in the middle, and the induced positive charge on the right. For decreasing separation at any given charge ratio, the enhancement in the induced negative charge in the medium region between the particles results in a further stabilization of the leftmost position of the induced negative charge $\widetilde{q}_{1}^{-}$ and the rightmost position of the induced positive charge $\widetilde{q}_{1}^{+}$ in the pictorial representation of sphere 1.

At very short interparticle separations, if the absolute value of the charge ratio is sufficiently large (e.g. $q_2/q_1 =16$ and $q_2/q_1 =80$), the induced negative charge in the medium region to the right of sphere 2 and the depletion-weakened induced negative charge in the medium region between the spheres together result in a repulsion of the induced negative charge $\widetilde{q}_{2}^{-}$ in sphere 2, from the rightmost position to the middle position in the corresponding pictorial representation.

~\\
\textbf{Scenario C: Opposite-charge attraction or repulsion in a more polarizable medium}
~\\

We have repeated our numerical investigations with a sign inversion of $q_2$ and with all other parameters remaining unchanged. The charge ratios considered are therefore (a) $q_2/q_1=-3$, (b) $q_2/q_1=-8$, (c) $q_2/q_1=-16$, and (d) $q_2/q_1=-80$, respectively. The results, which are summarized in Fig.~\ref{ksmalln}, are analogous to those for scenario B, where in both scenarios there is a crossover between attraction and repulsion. For decreasing interparticle separation, there is generally an increasing deviation of the electrostatic force $F_2$ on sphere 2 from the Coulomb force $F_\text{C}$, due to a strengthening of polarization-induced repulsion. At sufficiently large absolute values of $q_2/q_1$ (e.g. $q_2/q_1 =-8$, $q_2/q_1 =-16$ and $q_2/q_1 =-80$), the Coulomb attraction is overtaken by the polarization-induced repulsive force at short interparticle separations such that the overall interparticle interaction becomes counterintuitively repulsive.

For decreasing interparticle separation, the magnitude of induced charge in each sphere as relative to $q_{1}$, i.e. $\widetilde{q}_{i}/q_{1}$, generally increases. For the four different values of $q_2/q_1$, we observed qualitative differences in the variation of $\widetilde{q}_{i}/q_{1}$ for decreasing interparticle separation: (a) At $q_2/q_1 = -3$, neither the value of $\widetilde{q}_{1}/q_{1}$ nor that of $\widetilde{q}_{2}/q_{1}$ exceeds unity, and the interparticle interaction remains attractive all the way; (b) At $q_2/q_1 =-8$, the magnitudes of induced charge in both spheres are still all the way smaller than $q_1$, yet the value of $\widetilde{q}_{1}/q_{1}$ is not far from unity at short interparticle separations. This corresponds to a sufficiently strong effect of polarization for the occurrence of opposite-charge repulsion; (c) At $q_2/q_1 =-16$, the value of $\widetilde{q}_{1}/q_{1}$ exceeds unity at small interparticle separations but the value of $\widetilde{q}_{2}/q_{1}$ is still all the way smaller than unity, and opposite-charge repulsion occurs at short interparticle separations; (d) At $q_2/q_1 =-80$, both the value of $\widetilde{q}_{1}/q_{1}$ and that of $\widetilde{q}_{2}/q_{1}$ exceed unity at certain interparticle separations, and opposite-charge repulsion occurs at short interparticle separations. 

The positions of induced positive charge and induced negative charge in sphere 2 as relative to the sphere’s centre exhibit a non-trivial variation for decreasing interparticle separation, at all four values of $q_2/q_1$. While the positional order of charge in sphere 1 is the same for any interparticle separation or charge ratio, i.e. with the induced positive charge on the left, the free positive charge in the middle, and the induced negative charge on the right, the positional order of charge in sphere 2 exhibits an interesting dependence on the interparticle separation and on the charge ratio, as discussed below:

At (a) $q_2/q_1=-3$, the polarization of the medium region surrounding sphere 2 is relatively weak at large interparticle separations, such that the electrostatic force $F_2$ on sphere 2 is almost identical to the Coulomb force $F_\text{C}$. For decreasing interparticle separation, the induced positive charge in the medium region between the spheres results in an attraction of the induced negative charge $\widetilde{q}_{2}^{-}$ in sphere 2 from the rightmost position to the middle position in the corresponding pictorial representation. Despite such an alteration in charge distribution, the overall interparticle interaction is all the way attractive, because the attractive interaction between the induced negative charge $\widetilde{q}_{1}^{-}$ in sphere 1 and the induced positive charge $\widetilde{q}_{2}^{+}$ in sphere 2 remains dominant.

At (b) $q_2/q_1=-8$ and (c) $q_2/q_1=-16$, the polarization of the medium region surrounding sphere 2 is generally stronger than the case of $q_2/q_1=-3$. For decreasing interparticle separation, the induced positive charge in the medium region between the spheres results in an attraction of the induced negative charge $\widetilde{q}_{2}^{-}$ in sphere 2 from the rightmost position to the leftmost position in the corresponding pictorial representation. This results in a reversal of polarization in sphere 2 and a switch of the electrostatic force from attractive to repulsive, in which case the repulsive interaction between the induced negative charge $\widetilde{q}_{1}^{-}$ in sphere 1 and the induced negative charge $\widetilde{q}_{2}^{-}$ in sphere 2 becomes dominant.

At (d) $q_2/q_1 =-80$, the polarization of the medium region surrounding sphere 2 is stronger than the three cases described above. For decreasing interparticle separation, the induced positive charge in the medium region between the spheres not only results in an attraction of the induced negative charge $\widetilde{q}_{2}^{-}$ in sphere 2 from the rightmost position to the leftmost position in the corresponding pictorial representation, but also a repulsion of the induced positive charge $\widetilde{q}_{2}^{+}$ in sphere 2 from the leftmost position to the rightmost position. Like the cases of $q_2/q_1=-8$ and $q_2/q_1=-16$, this results in a reversal of polarization in sphere 2 and a switch of the electrostatic force from attractive to repulsive, in which case the repulsive interaction between the induced negative charge $\widetilde{q}_{1}^{-}$ in sphere 1 and the induced negative charge $\widetilde{q}_{2}^{-}$ in sphere 2 becomes dominant.

At very short interparticle separations, the induced positive charge in the medium region to the right of sphere 2 and the depletion-weakened induced positive charge in the medium region between the spheres together result in a repulsion of the induced positive charge $\widetilde{q}_{2}^{+}$ in sphere 2 from the rightmost position to the middle position in the corresponding pictorial representation.

If the dielectric constant of the surrounding medium is larger than the dielectric constants of two like-charged spheres, i.e. $k_1<k_0$ and $k_2<k_0$, the force $F_2$ acting on sphere 2 is an amplification of the Coulomb repulsion by a polarization-induced repulsive force. Else if the spheres are oppositely charged, the force $F_2$ experienced by sphere 2 is a mitigation of the Coulomb attraction by a polarization-induced repulsive force. As illustrated in Fig.~\ref{ksmallc}, a sign inversion of $q_2$ has little impact on the polarization-induced repulsive force. 

\section{Conclusions}\label{sec:conclusion}

By means of numerical simulations and theoretical analysis, we have investigated the mechanisms of electrostatic interactions between two charged dielectric spheres inside a polarizable medium. For short inter-particle separations with strong effects of polarization, our findings revealed a significant deviation of the interparticle force from that described by Coulomb's law and an intriguing correlation of the nature of interparticle interaction with the polarizability of the medium, as summarized in Fig. \ref{figclassification}. Wherever the force on the more polarizing particle switches from repulsive to attractive or vice versa for decreasing interparticle separation, the effective dipole in this particle undergoes a reversal in its orientation. Our findings, which have also revealed possible depletion effects of polarization charge in the medium for near-touching spheres, contribute to a comprehensive understanding of the mechanisms of electrostatic interactions between two charged dielectric spheres inside a polarizable medium, and have valuable implications for future explorations of multi-body systems.

For a pair of polarizable ions in a vacuum, the impossibility of opposite-charge repulsion has been derived analytically based on the empirical polarizability-volume relation of a polarizable-ion model \cite{PIchan2020theory}. As possible future work, for a system of two charged dielectric spheres, it might be possible to derive the impossibility of opposite-charge repulsion in a comparatively less polarizable medium and of like-charge attraction in a comparatively more polarizable medium if we are able to obtain an empirical relation between the effective-dipole moment and local electric field for each sphere.

\section{Conflicts of interest}

There are no conflicts of interest to declare.

\section{Acknowledgements}

ZG would like to acknowledge financial support from the Natural Science Foundation of China (Grant No. 12201146), the Natural Science Foundation of Guangdong Province (Grant No. 2023A1515012197), the Basic and Applied Basic Research Project of Guangzhou (Grant No.2023A04J0054), and the Guangzhou-HKUST(GZ) Joint Research Project (Grant Nos. 2023A03J0003 and 2024A03J0606). HKC would like to acknowledge financial support from the Natural Science Foundation of Guangdong Province (Grant no.: 2024A1515011585).

\providecommand*{\mcitethebibliography}{\thebibliography}
\csname @ifundefined\endcsname{endmcitethebibliography}
{\let\endmcitethebibliography\endthebibliography}{}


\end{document}